\documentclass[article]{aa}
\usepackage[varg]{txfonts}

\usepackage{ulem}
\usepackage{todonotes}

\usepackage{hyperref}
\hypersetup{colorlinks=true,linkcolor=blue,citecolor=blue,filecolor=blue,urlcolor=blue}

\usepackage{tikz,xcolor,graphicx}
\graphicspath{{images/}}

\usepackage{bm}

\definecolor{lime}{HTML}{A6CE39}
\DeclareRobustCommand{\orcidicon}{%
        \begin{tikzpicture}
        \draw[lime, fill=lime] (0,0)
        circle [radius=0.16]
        node[white] {{\fontfamily{qag}\selectfont \tiny ID}};
        \draw[white, fill=white] (-0.0625,0.095)
        circle [radius=0.007];
        \end{tikzpicture}
        \hspace{-2mm}
}

\newcommand{\orcidVP}{\href{https://orcid.org/0000-0002-3031-062X}{\orcidicon}}
\newcommand{\orcidMBD}{\href{https://orcid.org/0000-0001-6080-1190}{\orcidicon}}
\newcommand{\orcidHB}{\href{https://orcid.org/0000-0002-1959-6946}{\orcidicon}}
\newcommand{\orcidAB}{\href{https://orcid.org/0000-0002-4674-0704}{\orcidicon}}
\newcommand{\orcidICZ}{\href{https://orcid.org/0000-0001-9478-5731}{\orcidicon}}
\newcommand{\orcidMH}{\href{https://orcid.org/0000-0002-2363-5522}{\orcidicon}}
\newcommand{\orcidEL}{\href{https://orcid.org/0000-0001-7144-4766}{\orcidicon}}
\newcommand{\orcidAW}{\href{https://orcid.org/0000-0002-7501-9801}{\orcidicon}}

\newcommand{\tenc}{\tau_\mathrm{enc}}
\newcommand{\trh}[1][]{\tau_\mathrm{rh#1}}
\newcommand{\trc}[1][]{\tau_\mathrm{rc#1}}
\newcommand{\tgr}{\tau_\mathrm{gr}}
\newcommand{\tseg}{\tau_\mathrm{seg}}
\newcommand{\tej}{\tau_\mathrm{ej}}

\newcommand{\rh}[1][]{r_\mathrm{h#1}}
\newcommand{\Rh}[1][]{R_\mathrm{h#1}}
\newcommand{\Msun}{M_\odot}

\newcommand{\Rmin}{R_\mathrm{min}}
\newcommand{\Rlim}{R_\mathrm{lim}}
\newcommand{\abin}{a_\mathrm{bin}}
\newcommand{\Ebin}{E_\mathrm{bin}}
\newcommand{\Etot}[1][]{E_{\mathrm{tot}#1}}
\newcommand{\Nbin}{N_\mathrm{bin}}
\newcommand{\Nenc}[1][]{N_\mathrm{enc#1}}
\newcommand{\Mcl}{M_\mathrm{cl}}
\newcommand{\vbin}{v_\mathrm{bin}}
\newcommand{\vesc}{v_\mathrm{esc}}
\newcommand{\vinf}{v_\infty}

\newcommand{\Myr}{\mathrm{Myr}}
\newcommand{\Gyr}{\mathrm{Gyr}}
\newcommand{\kms}{\mathrm{km\,s^{-1}}}

\newcommand{\pc}{\mathrm{pc}}
\newcommand{\au}{\mathrm{au}}
\newcommand{\kpc}{\mathrm{kpc}}
\newcommand{\der}{\mathrm{d}}
\newcommand{\CRD}{\mathcal{C}}

\begin{document} 

\title{Spatial mixing of stellar populations in globular clusters via binary--single star scattering}
\subtitle{}
\titlerunning{Spatial mixing of stellar populations in globular clusters via binary--single star scattering}
\authorrunning{Pavlík et al.}

\author{%
    Václav Pavlík\inst{\ref{asu},\ref{iu},}\thanks{\email{pavlik@asu.cas.cz}}\orcidVP
    \and Melvyn B.~Davies\inst{\ref{lund}}\orcidMBD
    \and Ellen I.~Leitinger\inst{\ref{unibo},\ref{eso}}\orcidEL
    \and Holger Baumgardt\inst{\ref{uq}}\orcidHB
    \and Alexey Bobrick\inst{\ref{mon},\ref{ARC},\ref{tech}}\orcidAB
    \and Ivan Cabrera-Ziri\inst{\ref{vyoma}}\orcidICZ
    \and Michael Hilker\inst{\ref{eso}}\orcidMH
    \and Andrew J.~Winter\inst{\ref{mpia},\ref{oca}, \ref{qmul}}\orcidAW
}

\institute{%
    Astronomical Institute of the Czech Academy of Sciences, Bo\v{c}n\'i~II~1401, 141~00~Prague~4, Czech Republic \label{asu}
    \and Department of Astronomy, Indiana University, Swain Hall West, 727 E 3$^\text{rd}$ Street, Bloomington, IN 47405, USA \label{iu}
    \and Centre for Mathematical Sciences, Lund University, Box 118, SE-221 00 Lund, Sweden \label{lund}
    \and Dipartimento di Fisica e Astronomia, Università degli Studi di Bologna, Via Gobetti 93/2, I-40129 Bologna, Italy \label{unibo}
    \and European Southern Observatory, Karl-Schwarzschild-Str. 2, D-85748 Garching, Germany \label{eso}
    \and School of Mathematics and Physics, The University of Queensland, St. Lucia, QLD 4072, Australia \label{uq}
    \and School of Physics and Astronomy, Monash University, Clayton, Victoria 3800, Australia \label{mon}
    \and ARC Centre of Excellence for Gravitational Wave Discovery -- OzGrav, Australia \label{ARC}
    \and Technion -- Israel Institute of Technology, Physics Department, Haifa, Israel 32000 \label{tech}
    \and Vyoma GmbH, Karl-Theodor-Straße 55, 80803 Munich, Germany \label{vyoma}
    \and Max-Planck Institute for Astronomy (MPIA), Königstuhl 17, 69117 Heidelberg, Germany \label{mpia}
    \and Université Côte d'Azur, Observatoire de la Côte d'Azur, CNRS, Laboratoire Lagrange, 06300 Nice, France \label{oca}
    \and Astronomy Unit, School of Physics and Astronomy, Queen Mary University of London, London E1 4NS, UK \label{qmul} 
}

\date{Received August 5, 2025; accepted October 7, 2025}


\abstract
{The majority of Galactic globular star clusters (GCs) have been reported to contain at least two populations of stars (hereafter, we use P1 for the primordial and P2 for the chemically-enriched population). Recent observational studies found that dynamically-old GCs have P1 and P2 spatially mixed due to relaxation processes. However, in dynamically-young GCs, where P2 is expected to be more centrally concentrated from birth, the spatial distributions of P1 and P2 are sometimes very different from system to system. This suggests that more complex dynamical processes specific to certain GCs might have shaped those distributions.}
{We aim to investigate the discrepancies between the spatial concentration of P1 and P2 stars in dynamically-young GCs. Our main focus is to evaluate whether massive binary stars (e.g.~black holes) can cause the expansion of the P2 stars through binary--single interactions in the core, and whether they can mix or even radially invert the P1 and P2 distributions.}
{We use a set of theoretical and empirical arguments to evaluate the effectiveness of binary--single star scattering. We then construct a set of direct $N$-body models with massive primordial binaries to verify our estimates further and gain more insights into the dynamical processes in GCs.}
{We find that binary--single star scatterings can push the central P2 stars outwards within a few relaxation times. While we do not produce radial inversion of P1 and P2 for any initial conditions we tested, this mechanism systematically produces clusters where P1 and P2 look fully mixed even in projection. The mixing is enhanced 1) in denser GCs, 2) in GCs containing more binary stars, and 3) when the mass ratio between the binary components and the cluster members is higher.}
{Binary--single star interactions seem able to explain the observable properties of some dynamically-young GCs (e.g.~NGC~4590 or NGC~5904) where P1 and P2 are fully radially mixed.}


\maketitle
%

\section{Introduction}
\label{sec:introduction}

Star clusters are often used as laboratories for studying star formation, stellar evolution and the dynamics of galaxies. When focusing on the old ($>2\,\Gyr$) and relatively massive ($>10^4\,\Msun$) globular clusters (GCs), these gravitationally bound systems have long been known to host multiple populations of stars with distinct chemical compositions and evolutionary histories
\citep[see, e.g.][and references therein]{carretta_etal2010, gratton_review2012, hst_UV_legacy, multipop_review18, martocchia2018, hst_legacy_xviii, Tio_Ves_Var19, martocchia2020PhD, sollima21, vesperini_etal21}.
Hereafter, stars from the primordial (or first) population are referred to as `P1', and the enriched second population as `P2'.
For instance, variations in the abundances of light elements point to complex formation scenarios of P1 and P2, including the accretion of enriched material onto existing stars or a direct formation of P2 stars out of this enriched material \citep[e.g.][]{carretta_etal2010, bekki_2011}.
In some clusters, P1 and P2 are also characterised by different dynamical evolution, which leads to measurable differences in present-day kinematic properties, such as rotation or anisotropy in velocity dispersion \citep{cordoni_etal20a, cordoni_etal20b, dalessandro_etal24, leitinger_etal24, cadelano_etal2024}. Since velocity anisotropy affects the relaxation time scales, mass segregation, and the evolution towards energy equipartition \citep[see, e.g.][]{Tio_Ves_Var16, cluster_history, pav_ves_letter, pav_ves2, pav_ves3, pavlik_etal_tangential2024}, it is expected that each stellar population might evolve differently \citep[as shown, e.g.\ by][]{livernois_etal24}.

Most models of GC formation predict that P2 stars are being preferentially born in their cores \citep[e.g.][]{dercole_etal2008, lacchin_etal22, yaghoobi_etal22a, yaghoobi_etal22b}.
Although the spatial distribution of these populations evolves due to relaxation processes or gas expulsion \citep[e.g.][]{vesperini_etal2013, decressin_etal2010}, dynamically younger GCs are expected to retain their initial structural differences \citep[e.g.][and references therein]{dalessandro_etal19, dalessandro_etal24}.
However, a recent observational study by \citet{leitinger_etal23} showed that the spatial distribution of P1 and P2 varies among the dynamically-young Galactic GCs.
Notably, while P2 stars in some GCs (e.g.\ NGC~5024 or NGC~6809) appear to be more centrally concentrated than P1 (as expected for GCs less than a few relaxation times old), in other clusters which have similar dynamical ages as these ones (e.g.\ NGC~3201 or NGC~6101), P2 stars primarily occupy the outer regions of the clusters and P1 is preferentially in the central regions.
There are also several dynamically-young GCs which are fully mixed, i.e.\ they show no preferred concentration of either P1 or P2 (e.g.\ NGC~288, NGC~4590, NGC~5053, NGC~5904, NGC~6205, NGC~7078, and NGC~7089).
At first glance, these findings contradict most models of GC formation that rely on P2 stars being born in a centrally concentrated manner.
A possible explanation for this variety of morphologies may, however, also be that the internal dynamical processes (which might play out differently in each GC) are responsible for an early population mixing or inversion of an initially P2-centrally-concentrated GC, to instead show P2 stars in the outer regions.

We investigate whether the interactions between binary and single stars may be the dynamical origin of population mixing and inversion in GCs. Our preliminary study \citep{modest25_poster} with several small and compact clusters supported this idea. However, it only focused on the idealised case with full tangential stellar velocity anisotropy, where any systematic radial motion early in the cluster evolution would be produced by binary--single star scattering in the cluster core. Here we investigate a larger suite of more realistic initial conditions and also elucidate the effects theoretically. A similar dynamical scenario, where stars are heated by a black hole subsystem, has also been independently proposed in a recent study by \citet{mpop_mocca_dynamics}.

Generally, the binary--single star scattering is a prevalent cause of stellar ejections from the GC core \citep[see, e.g.][]{aarseth1972, oleary, heggie_hut} and is usually mostly visible when the GC goes through core collapse later in its dynamical evolution \citep[see, e.g.][]{tanikawa, tanikawa2, fujii_pz, pavl_subr}. However, suppose a GC contains massive stars initially (O or B-type of several tens of $\Msun$). In that case, they will most likely be paired in binaries \citep{kroupa95a, goodwin_kroupa05} and positioned in the core very early on -- either the GC is already formed mass segregated, as has been shown for young star clusters \citep[][]{plunkett, pav_segr, pav_binseg} but also for older GCs \citep{zonoozi_etal24}, or they quickly sink into the core due to mass segregation \citep[e.g.][]{binney_tremaine}.
Some of those very massive stars would evolve into black holes (BHs) on a time scale of a few Myr \citep[e.g.][]{sse, bse, belczynski_etal08, belczynski_etal10, ziosi_etal14}.
If their supernova natal kick velocity is then adequately small \citep[see, e.g.][]{lyn94,podsialdowski_etal04}, they may be retained in the GC \citep{peuten_etal16, banerjee_gw17, baumgardt_sollima17, pavl_bh}.
Hence, BH--BH binaries might be produced promptly in GCs, subsequently ejecting P2 stars from the core to the outer regions through binary--single star scattering. 

Our paper is organised as follows: in Section~\ref{sec:theory}, we present theoretical estimates for the binary--single star scattering, in Section~\ref{sec:models}, we show numerical simulations designed to test our theory, in Section~\ref{sec:results}, we focus on the evolution of our models, in Section~\ref{sec:observations}, we compare our key findings with real observed GCs, in Section~\ref{sec:discussion}, we discuss our results and possible alternative scenarios, and we summarise our conclusions in Section~\ref{sec:conclusions}.

\section{Theoretical estimates}
\label{sec:theory}

We are interested in understanding the degree to which the binary--single scattering can eject an initially centrally concentrated P2 stellar population to the outer regions of the GC. In this section, we will estimate the rate of encounters and the typical properties of such binary and single stars.
There are three important time scales to consider:
1) the time scale for a given binary to encounter a random single star, likely leading to the ejection of the single star from the core;
2) the time scale on which a significant fraction of the stars within the core encounter a binary and are, therefore, removed from the core;
and 3) the time scale on which stars from outside the core will sink into the core.
As the core originally contains only P2 stars with the P1 stars outside of the core, an inversion of the stellar population (i.e.~P2 ending up largely placed further out than P1) will occur if the second time scale is shorter than the third but also that (for some reason) the binary encounters cease
after this time.
If alternatively, the third time scale is shorter than the second, and/or the binary-single scattering process continues, then instead the P1 and P2 populations will be mixed by binary--single scattering.

In a first approximation, we consider equal-mass single stars in a GC (with masses $m_3$) and a binary star with components $m_1$ and $m_2$ (typically two BHs), such that $(m_1 + m_2) \gg m_3$. Then the gravitationally focused close encounter timescale is \citep[see, e.g.~Eq.~7.196 in][]{binney_tremaine}
\begin{equation}
    \label{eq:tenc}
    \tenc \approx 32.5\,\Myr \ 
        \bigg(\frac{10^5 / \pc^3}{n}\bigg)
        \bigg(\frac{\Msun}{M}\bigg)
        \bigg(\frac{10\,\au}{\Rmin}\bigg)
        \bigg(\frac{\vinf}{10\,\kms}\bigg) \,,
\end{equation}
where $n$ is the stellar density ($10^4$ to $10^5\,\pc^{-3}$ for the central regions of a typical GC),
$M$ is the total mass of the interacting bodies (i.e.~$M = m_1 + m_2 + m_3$),
$\Rmin$ is the distance of the closest approach (where strong scattering typically occurs when $\Rmin\approx\abin$, the binary semi-major axis),
and $\vinf$ is the stellar velocity at infinity (comparable to the velocity dispersion of the $m_3$ stars).
Thus, the question we aim to answer is, what are the possible parameters for the binary to have a sufficiently short timescale $\tenc$ to produce enough scatterings while the cluster is dynamically young?

The energy transfer scales with the binary properties as follows. The binary binding energy is
\begin{equation}
    \label{eq:Ebin}
    \Ebin = \frac{G m_1 m_2}{2 \abin} \,,
\end{equation}
where $G$ is the Newtonian gravitational constant.
The energy of a hard binary is $\Ebin > \frac{1}{2} m_3 \vinf^2$ \citep{heggie75, hut, heggie_hut} for the $m_3$ perturber mass.
Or in terms of the semi-major axis, the hard/soft limit is
\begin{equation}
    \label{eq:abin_hs}
    \abin \approx 8.87\,\au \
        \left(\frac{m_1 \, m_2}{m_3\,\Msun}\right)
        \left(\frac{10\,\kms}{\vinf}\right)^2 \,
\end{equation}
where we used $G \approx 887 \,\au\,\left(\kms\right)^2\,\Msun^{-1}$.
The change in binding energy after an encounter with an incoming single star is
\begin{equation}
    \label{eq:DEbin1}
    \Delta\Ebin = \frac{1}{2} (m_1 + m_2) \vbin^2 + \frac{1}{2} m_3 v_3^2 \,,
\end{equation}
where $\vbin$ and $v_3$ are the typical velocities per encounter between the binary's centre of mass and a single star in the GC, respectively.
Using the conservation of momentum in the binary's centre-of-mass frame, i.e.\ $(m_1+m_2) \vbin = m_3 v_3$, Eq.~\eqref{eq:DEbin1} becomes
\begin{equation}
    \label{eq:DEbin2}
    \Delta\Ebin = \frac{1}{2} \frac{m_3 (m_1 + m_2 + m_3)}{m_1+ m_2} v_3^2 \,.
\end{equation}
The average change of binding energy during an encounter can be rewritten as a fraction of the binary binding energy, i.e.\ $\Delta\Ebin = \alpha\Ebin$, where $\alpha$ is a distribution function describing the strength of the encounters.
The mean stellar speed per encounter then becomes
\begin{equation}
    \label{eq:v3}
    v_3 = \sqrt{\alpha \, G \, \frac{m_1 m_2(m_1 + m_2)}{m_3(m_1 + m_2 + m_3)} \, \frac{1}{\abin}} \,.
\end{equation}

\begin{figure}
    \centering
    \includegraphics[width=\linewidth]{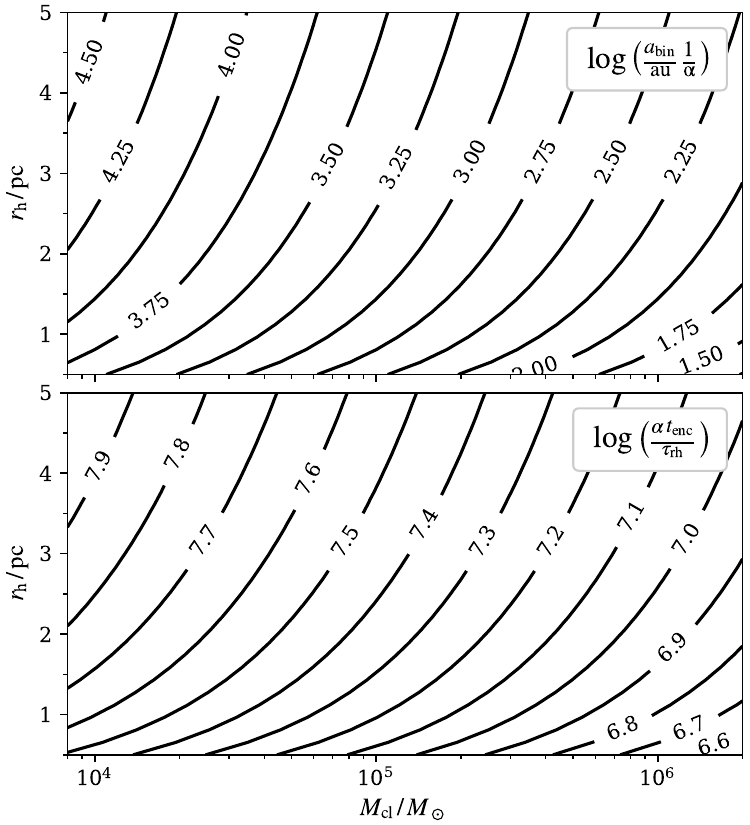}
    \caption{Estimated semi-major axis (top) and the encounter time scale (bottom) of a binary star that can eject single stars from the core in a cluster of a given mass and half-mass radius. In this plot, we used Eqs.~\eqref{eq:abin_alpha} and~\eqref{eq:tenc_alpha}, for clusters containing stars of equal masses ($m_3=1\,\Msun$), one binary star with $m_1=m_2=30\,\Msun$ at its centre, and $\vinf = 10\,\kms$. Time in the bottom panel is normalised by the half-mass relaxation time, Eq.~\eqref{eq:trh}.}
    \label{fig:estimates}
\end{figure}

Assuming, for instance, that the cluster follows a Plummer model \citep{plummer}, the stellar escape velocity from radius $r$ is
\begin{equation}
    \label{eq:vesc}
    \vesc(r)
        \approx \sqrt{2 \, G \Mcl} \left[r^2 + \left(\frac{\rh}{1.3}\right)^2\right]^{-1/4} \,,
\end{equation}
where $\rh$ is the half-mass radius and $\Mcl$ is the cluster mass, i.e.\ $\Mcl = Nm_3 + \Nbin(m_1+m_2) \approx Nm_3$ (if the mass of the binary stars is negligible compared to the cluster mass). Thus, to eject single stars from the core region, i.e.~\hbox{$v_3 \gtrsim \vesc(0)$}, Eqs.~\eqref{eq:v3} and~\eqref{eq:vesc} give the required binary star semi-major axis
\begin{equation}
    \label{eq:abin_alpha}
    \abin \lesssim 0.38 \, \alpha \, \rh \, \frac{m_1 m_2 (m_1 + m_2)}{m_3 (m_1 + m_2 + m_3)} \frac{1}{\Mcl} \,.
\end{equation}
Using Eqs.~\eqref{eq:tenc} and~\eqref{eq:abin_alpha} and taking $\Rmin=\abin$, the time scale for this binary to encounter a random single star (and get it ejected) is
\begin{equation}
    \label{eq:tenc_alpha}
    \tenc \gtrsim \frac{1.26 \cdot 10^3 \,\Myr}{\alpha}\
        \bigg(\frac{\rh}{\pc}\bigg)^2
        \bigg(\frac{m_3^2\ \Msun}{m_1 m_2 (m_1 + m_2)}\bigg)
        \bigg(\frac{\vinf}{10\,\kms}\bigg) \,.
\end{equation}
For an example of a BH--BH binary ($m_1 = m_2 = 30\,\Msun$) and a single star ($m_3 = 1\,\Msun$)
in a cluster with $\rh=0.5\,\pc$ and \hbox{$\vinf=10\,\kms$}, Eq.~\eqref{eq:tenc_alpha} yields $\tenc \gtrsim \alpha^{-1} \, 5.83 \cdot 10^{-3} \,\Myr$. We also plot other estimates derived from Eqs.~\eqref{eq:abin_alpha} and~\eqref{eq:tenc_alpha} in Fig.~\ref{fig:estimates}.

To achieve mixing (or radial inversion) of P1 and P2, we need to remove a significant fraction of P2 stars (i.e.\,$\Nenc[P2] \sim N_{\rm P2}$) from the core where they were born. Therefore, the time scale for a binary star to encounter and eject most/all of them is
\begin{equation}
    \label{eq:tej}
    \tej \equiv \tenc \Nenc[P2] \,.
\end{equation}
However, a single binary can only interact with a fraction of nearby stars whose trajectories are within its cross-section. Thus, to allow most stars in the core to have the opportunity to interact with the binary over $\tej$, their orbits must be sufficiently perturbed by relaxation. This implies that the time scales are $\trc \lesssim \tej$, where $\trc$ is the core relaxation time \citep[see, e.g.][]{heggie_hut}, defined as 
\begin{equation}
    \label{eq:trc}
    \trc(t) = \frac{0.34 \, \sigma^3_{\rm c}(t)}{\langle m_{\rm c}(t) \rangle \, \rho_{\rm c}(t) \, \ln{[0.4N(t)]}} \,,
\end{equation}
where
$\sigma_{\rm c}$ is the core velocity dispersion,
$\langle m_{\rm c} \rangle$ is the mean stellar mass in the core,
$\rho_{\rm c}$ is the core mass-density,
and $N(t)$ is the number of stars in the cluster at time $t$.\!\footnote{
We note that the core relaxation time may be slightly different depending on the stellar density in the core and the velocity distribution \citep[see, e.g.][]{binney_tremaine, merritt_dynamics, pavlik_etal_tangential2024}.}

To observe a mixed/inverted dynamically-young GC, most of the binary--P2-star encounters need to occur early in the cluster evolution, so $\tej$ per binary must be much lower than the half-mass relaxation time of the cluster \citep[e.g.][]{spitzer_hart_relax}
\begin{equation}
    \label{eq:trh}
    \trh(t) \approx \frac{0.138}{\ln{[0.4 N(t)]}} \sqrt{\frac{N(t)\, \rh^3(t)}{G\, m_3}} \,.
\end{equation}
This is because even if we start from completely P2-centrally-concentrated initial conditions, the P1 stars outside of the core will fall inwards as the cluster relaxes.
Thus, we should expect the number of binary--P2-star scatterings to reduce over time and the number of binary--P1-star scatterings to increase. Once the probability of scattering P1 and P2 becomes comparable, the binary would only be able to achieve mixing of the two populations because P1 will get scattered just as much as P2.
Consequently, to achieve inversion, the majority of P2 ejections via binary--single scattering need to happen early on and then stop.

\begin{figure*}
    \centering
    \includegraphics[width=0.56\linewidth]{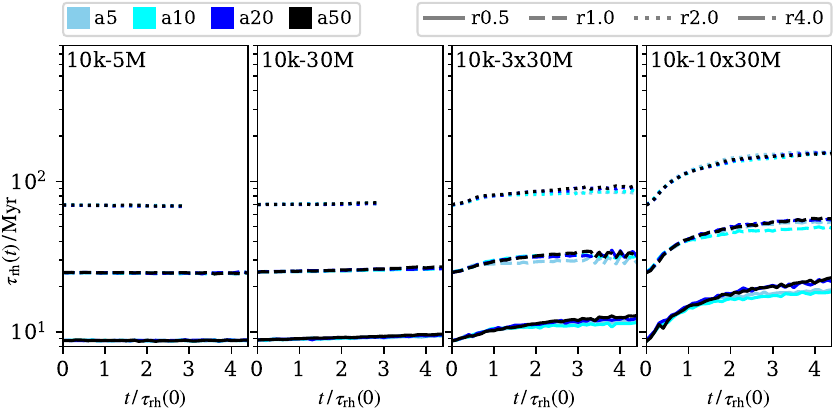}
    \hfill
    \includegraphics[width=0.427\linewidth]{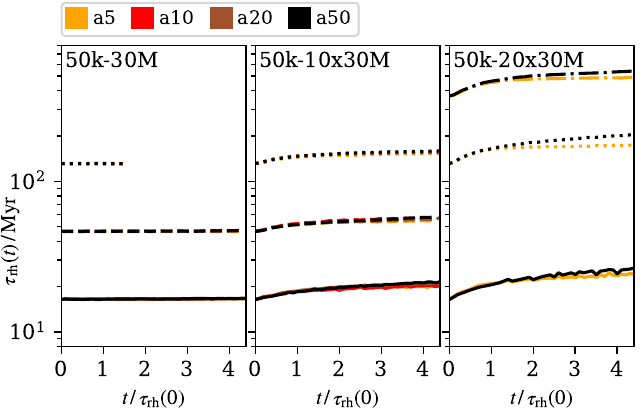}
    \caption{Evolution of the half-mass relaxation time, $\trh(t)$, calculated from Eq.~\eqref{eq:trh} in terms of the initial half-mass relaxation time, $\trh(0)$. The left-hand set of panels is for the \texttt{10k-*} clusters, the right-hand set is for \texttt{50k-*}. The models are separated by the primordial binary star numbers and masses (columns), the binary star initial semi-major axes (colours), and the initial cluster half-mass radius (line styles). Each line is averaged over all realisations of the corresponding model. (We note that the models \texttt{10k-5M*} and \texttt{50k-30M*} do not show any differences from the models without binaries, which we also integrated for comparison but do not display them here.)}
    \label{fig:Trh}

    \vspace{\floatsep}
    
    \includegraphics[width=0.56\linewidth]{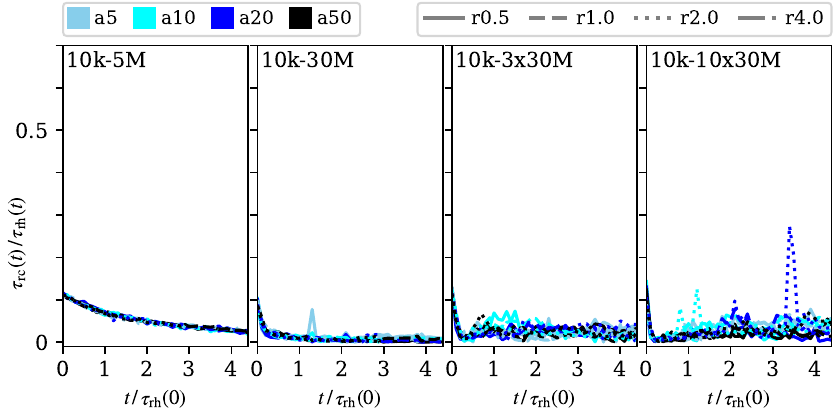}
    \hfill
    \includegraphics[width=0.427\linewidth]{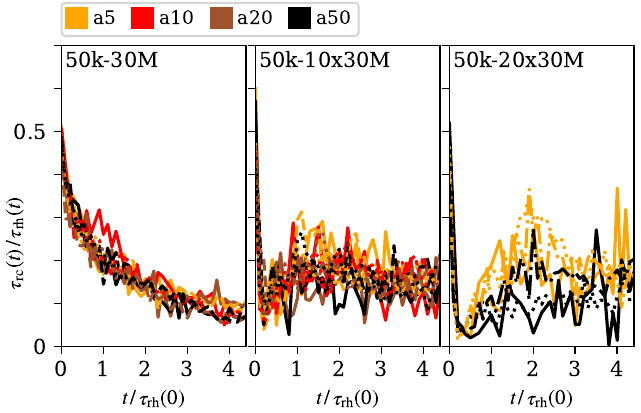}
    \caption{Evolution of the core relaxation time, $\trc(t)$ in the units of $\trh(t)$, calculated from Eq.~\eqref{eq:trc}. The horizontal axis shows time in terms of the initial half-mass relaxation time, $\trh(0)$. The left-hand set of panels is for the \texttt{10k-*} clusters, the right-hand set is for \texttt{50k-*}. The models are separated by the primordial binary star numbers and masses (columns), the binary star initial semi-major axes (colours), and the initial cluster half-mass radius (line styles). Each line is averaged over all realisations of the corresponding model.}
    \label{fig:Trc}
\end{figure*}

We must explore how to set these three timescales above to kick most of the P2 outwards very early.
We note that in this section, we have assumed that the single $m_3$-stars must escape from the GC completely. However, \citet{leitinger_etal23} have shown that the radial distribution of P2 in some GCs (e.g.~NGC~6101 and NGC~3201) reaches its maximum at the projected half-light radius.\!\footnote{The half-light and half-mass radii coincide in an equal-mass cluster.} Thus, mixing and partial radial inversion of P1 and P2 may also be driven by slightly less energetic encounters, and, for instance, the values of the semi-major axis in Fig.~\ref{fig:estimates} and Eq.~\eqref{eq:abin_alpha} are rather the lower estimates.

\section{Numerical models}
\label{sec:models}

\subsection{Initial conditions of the star clusters}

We performed a grid of simulations to test the theoretical estimates above.
We use clusters with $N = 10^4$ stars (to probe the parameter space with computational simplicity) and $N = 5{\cdot}10^4$ stars (for a more realistically sized system). Some of these stars are paired into $\Nbin = 1$, 3, 10 or 20 massive primordial binaries, and the remaining ones are single equal-mass stars (with $m_3 = 1\,\Msun$).
The clusters are isolated, have a \citet{plummer} profile with the half-mass radius $\rh = 0.5$, $1.0$, $2.0$ or $4.0\,\pc$, isotropic velocity distribution, and are initialised in virial equilibrium.
The initial conditions were generated with \textsc{McLuster} \citep{mcluster} and integrated with the $N$-body tree code \textsc{PeTar} \citep{petar} in several random realisations each.\!\footnote{Ten realisations of each smaller model and five of the larger models.} The simulation time corresponds to several $\trh(0)$ -- see also Fig.~\ref{fig:Trh} with the relaxation times of the models for comparison.
We note that in Section~\ref{sec:observations}, we compare these models with significantly older GCs in terms of physical units of time (age of several Gyr, whereas our models range between 100\,Myr to 2.5\,Gyr), however, in terms of their dynamical age (i.e.~the number of $\trh$), the observed systems are similar to our models.

In each model, we split the single $m_3$-stars into two populations based on their total initial energy, where for the $i$-th star
\begin{equation}
    \label{eq:Etot}
    \Etot[,i] =
    \frac{1}{2} m_i v_i^2
    - G \Mcl\,\left[r_i^2 + \left(\frac{\rh}{1.3}\right)^2\right]^{-1/2} \,.
\end{equation}
We label as `P2' the stars with total energy in the lowest quartile, and the rest are labelled `P1'.\!\footnote{We note that the primordial binaries are not counted towards either population, but they do contribute to the potential term in Eq.~\eqref{eq:Etot}.}
This separation is based on the assumption that P2 stars are born more concentrated and would have their orbits initially more confined in the central regions. It is usually assumed that P2 would be formed 30--100\,Myr after P1 if the cause of enrichment is winds from the asymptotic-giant-branch stars or from interacting binaries \citep[e.g.][]{multipop_review18}. However, in other formation scenarios, where the enrichment is caused by very-/super-massive stars \citep[e.g.][]{gieles_etal2018} or stellar-wind mass-loss from stellar merger events \citep[e.g.][]{wang_etal2020}, P2 may form within a few Myr after P1.
Simulating two populations formed at two different times is beyond the scope of this paper; nevertheless, even our simple distinction between P1 and P2 helps us reveal several effects of the binary--single star interactions and spatial redistribution of the stars.

\subsection{Primordial binary stars}

The binary components have $m_1 = m_2 = 5$ or $30\,\Msun$ depending on the model.\!\footnote{We note that these are typical BH masses and that some detected gravitational-wave sources even exceed $30\,\Msun$. The upper limit of all massive objects per cluster also does not exceed the constraints coming from more realistic systems. Specifically, applying stellar evolution \citep[from][]{sse,bse} on the \citet{kroupa} initial mass function (from $0.08$ to $150\,\Msun$) results in ${\approx}6\,\%$ of the initial total GC mass being contained in BH progenitors.}
They are generated on circular orbits with semi-major axis $\abin = 5$, 10, 20, or $50\,\au$ -- this makes them very hard.\!\footnote{See Eq.~\eqref{eq:abin_hs} where the theoretical hard/soft limit is $a \approx 200\,\au$ for a $(5{+}5)\,\Msun$ binary and ${\approx}8000\,\au$ for a $(30{+}30)\,\Msun$ binary.}
Even the choice of the widest $50\,\au$ binary is substantially small to theoretically drive the radial redistribution of P1 and P2 stars (see Fig.~\ref{fig:estimates}).

The binaries are initially randomly placed inside the central region, but they quickly segregate to the core, from where they start ejecting stars. The mass segregation time scale may be estimated from \citep[][]{spitzer_hart_relax, spitzer, binney_tremaine}
\begin{equation}
    \label{eq:tseg}
    \tseg \approx \frac{m_3}{m_1} \trh \,,
\end{equation}
where we approximate the average stellar mass by $m_3$.
Plugging in the numbers from each model and using the relaxation time from Fig.~\ref{fig:Trh}, we may verify that $\tenc \gg \tseg$, hence the binary does not have much time to interact with the $m_3$-stars while sinking towards the GC core.

The names of the models in this paper follow the convention:
$$[N] \ \texttt{k-} \
[\Nbin\,\texttt{x}\,({\rm if}\ {>}1)] \
[m_1/\Msun] \ \texttt{M} \
\texttt{r} \ [\rh/\pc] \
\texttt{a} \ [\abin / \au]$$
For instance a model with $10^4$ stars, $\rh=1\,\pc$, one binary of $(5{+}5)\,\Msun$ and $\abin=20\,\au$ is called \texttt{10k-5Mr1.0a20}.
We also use a wildcard `\texttt{*}' to refer to a group of models -- e.g.\ all models with $10^4$ stars, three $(30{+}30)\,\Msun$ binaries and $\rh = 0.5\,\pc$, regardless of the binary's semi-major axis, are collectively called \texttt{10k-3x30Mr0.5*}.
We also note that we did not follow all possible combinations of the initial conditions due to the cost of the simulations and their predictive nature.

\begin{figure*}
    \centering
    \includegraphics[width=\linewidth]{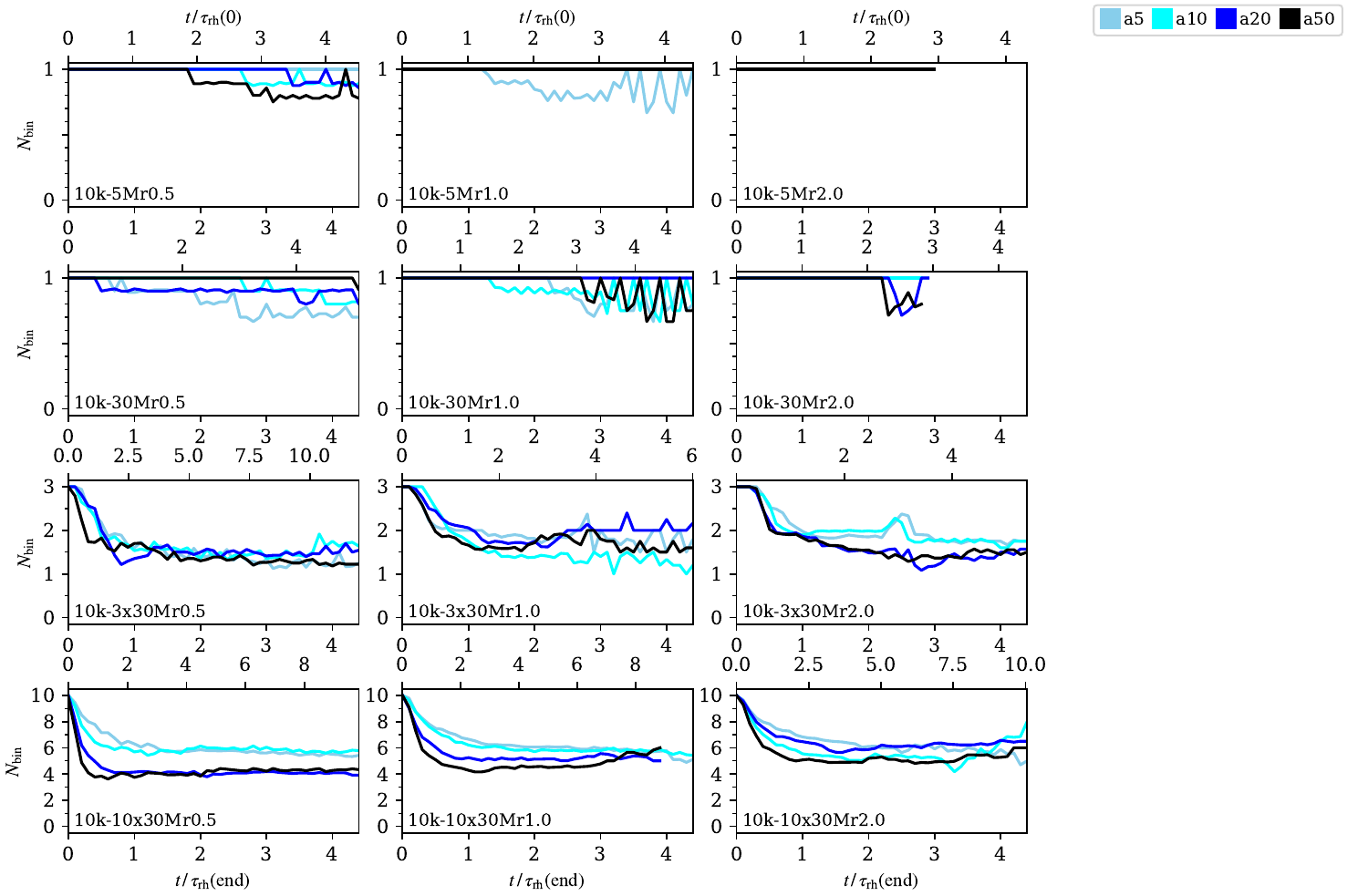}
    \caption{The number of massive binaries in each \texttt{10k-*} model in time (averaged over each model's realisations).
    The models are separated by their initial $\rh$ (columns), primordial binary systems (rows), and their semi-major axes (colours).
    In each plot, the top horizontal axis shows the evolutionary time in the initial half-mass relaxation times, and the lower horizontal axis is in multiples of $\trh$ at the end of the simulation (to mimic the `present-day' value).
    While we permit the dynamical formation of the low-mass ($m_3{+}m_3$) or mixed ($m_{1,2}{+}m_3$) binaries in the simulations, the numbers here only show the massive ($m_1{+}m_2$) binaries, including component switching. However, if the primordial binary goes through a state of an unstable triple, it is not counted towards $\Nbin$ (which is why we sometimes see the number of binaries decreasing and then increasing again).}
    \label{fig:Nbin_10k}
\vspace{\floatsep} 
    \centering
    \includegraphics[width=\linewidth]{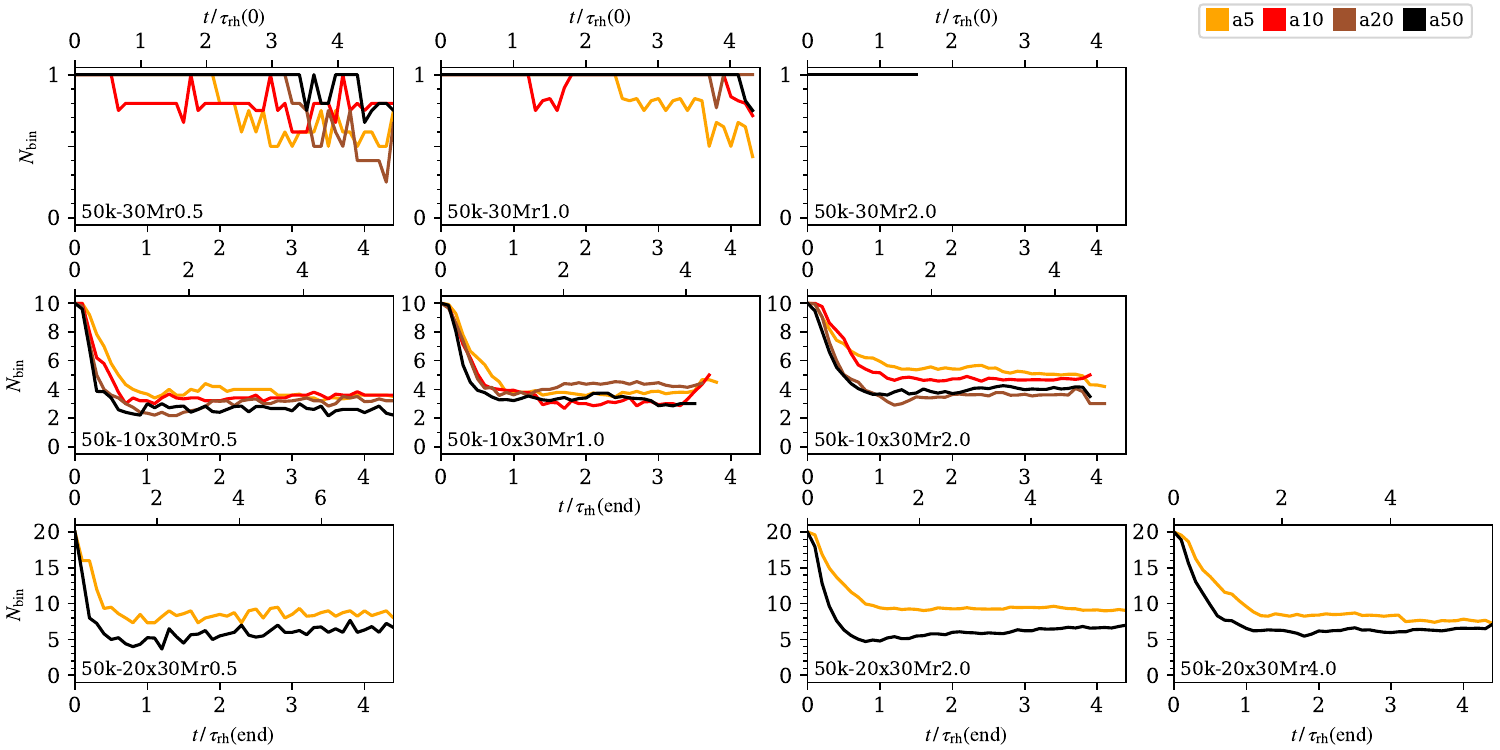}
    \caption{Same as Fig.~\ref{fig:Nbin_10k} but for the \texttt{50k-*} models.}
    \label{fig:Nbin_50k}
\end{figure*}

\begin{figure*}
    \centering
    \includegraphics[width=\linewidth]{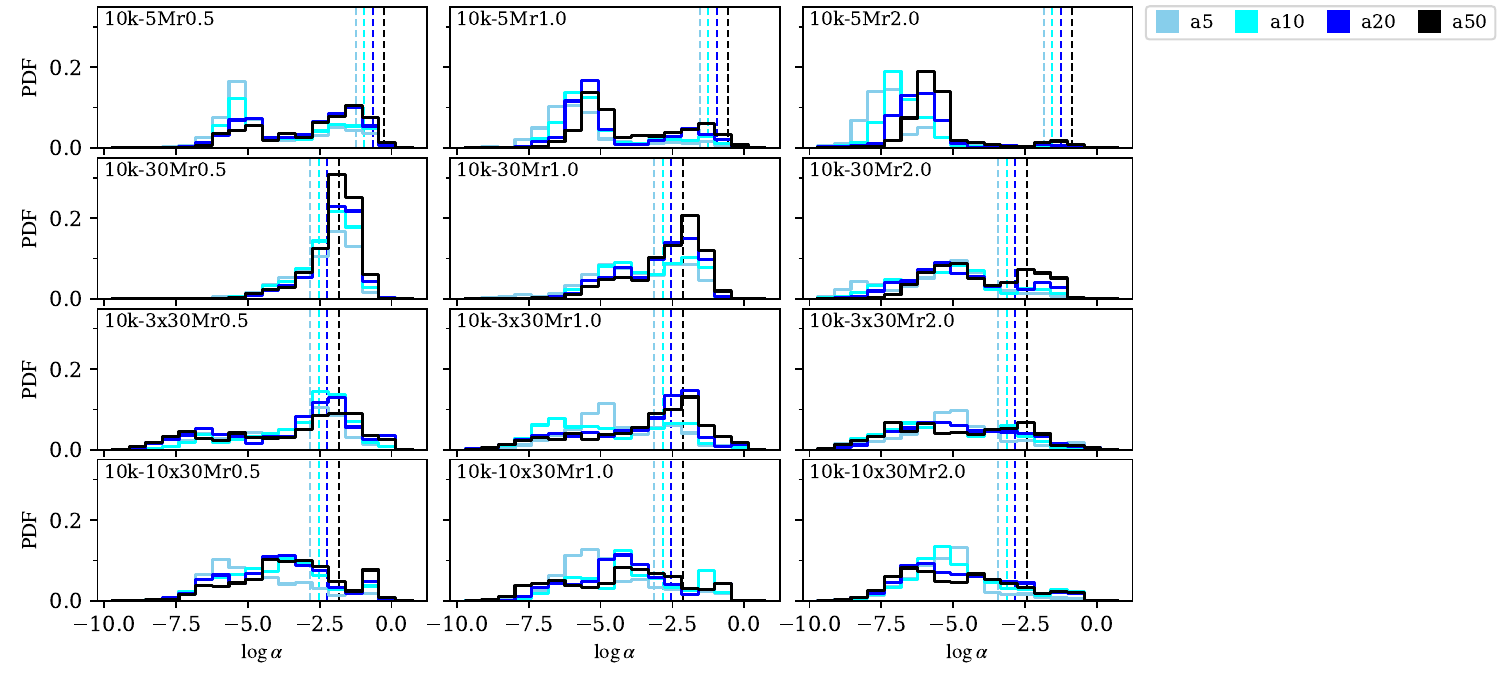}
    \caption{The probability density function (PDF) of the binding energy change, $\alpha$, per interaction, in the \texttt{10k-*} models. Binary--single and binary--binary interactions are counted together. For reference, we also plot with dashed vertical lines the theoretical values of $\alpha$, for which a single $m_3$-star should theoretically escape, as derived using Eq.~\eqref{eq:abin_alpha}.
    The figure is separated by each model's initial half-mass radius (columns), number and masses of the primordial binaries (rows), and the binary semi-major axes (colours).}
    \label{fig:alpha_10k}
\vspace{\floatsep} 
    \centering
    \includegraphics[width=\linewidth]{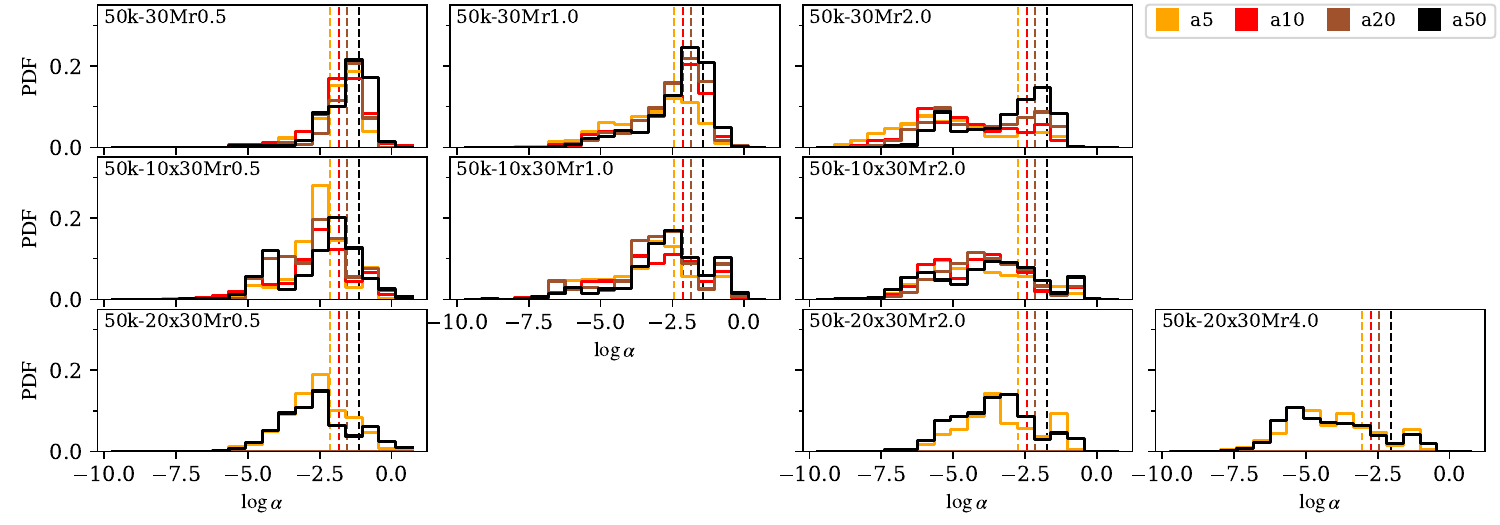}
    \caption{Same as Fig.~\ref{fig:alpha_10k} but for the \texttt{50k-*} models.}
    \label{fig:alpha_50k}
\end{figure*}

\begin{figure*}
    \centering
    \includegraphics[width=0.49\linewidth]{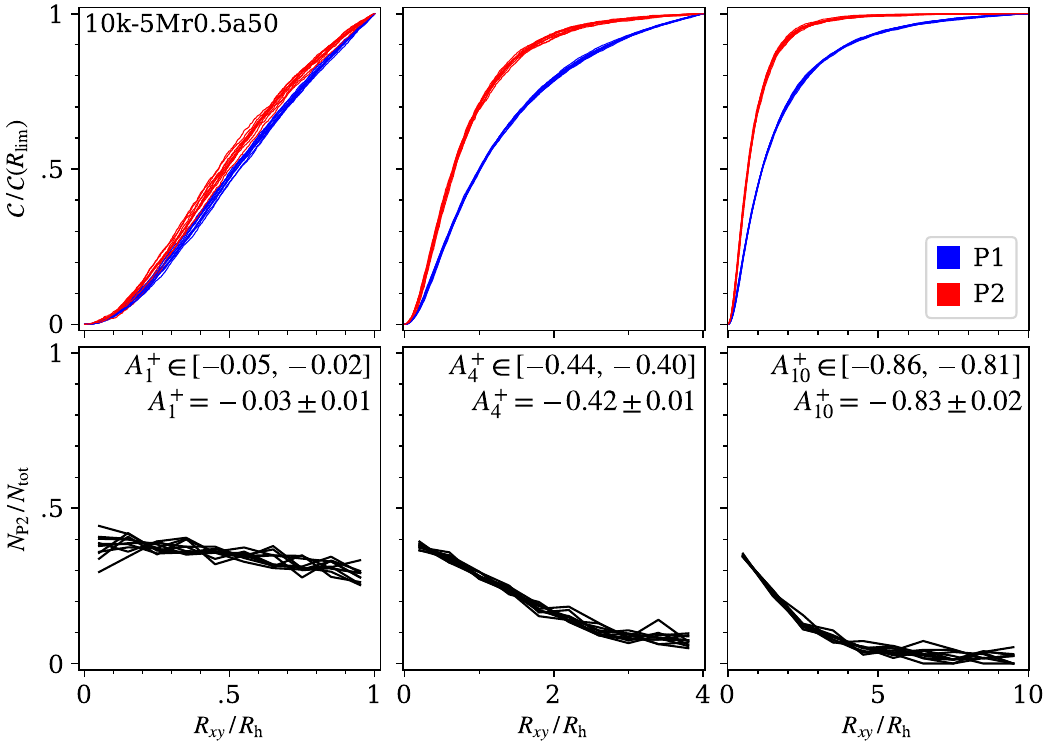}
    \hfill
    \includegraphics[width=0.49\linewidth]{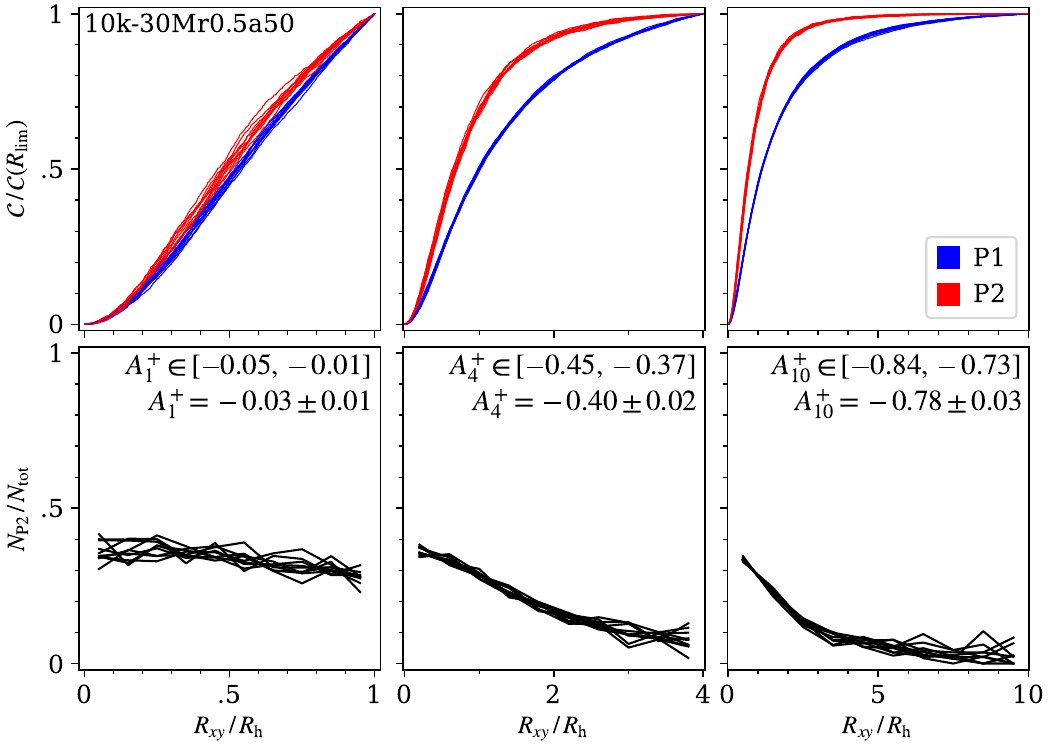}\\
    \includegraphics[width=0.49\linewidth]{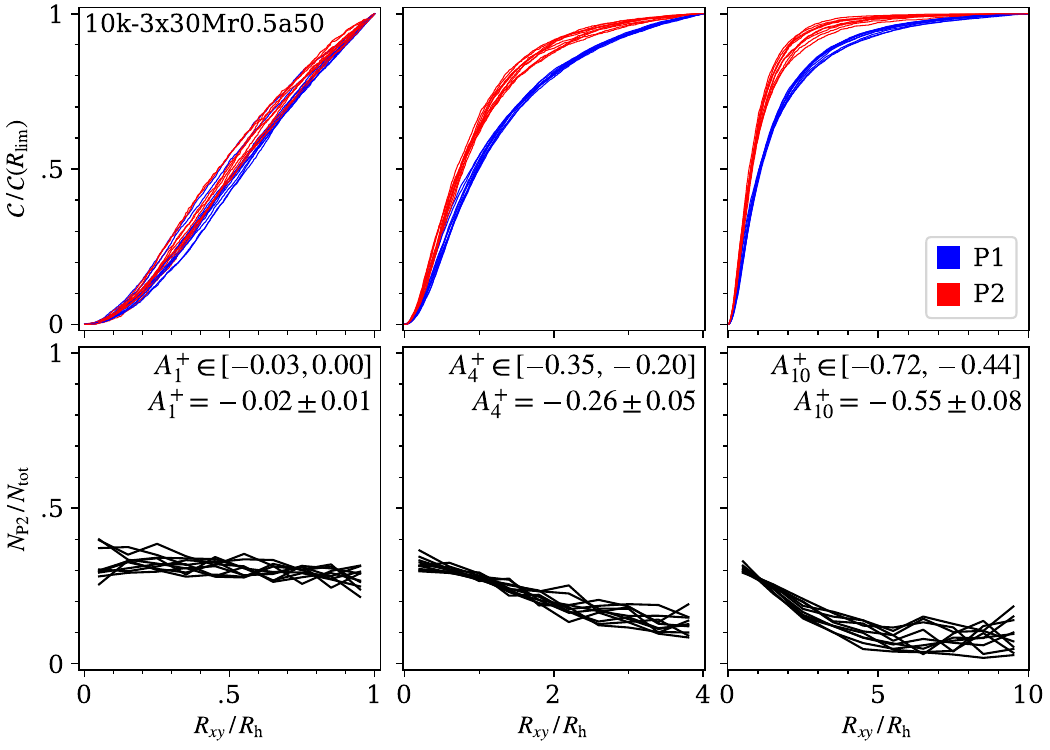}
    \hfill
    \includegraphics[width=0.49\linewidth]{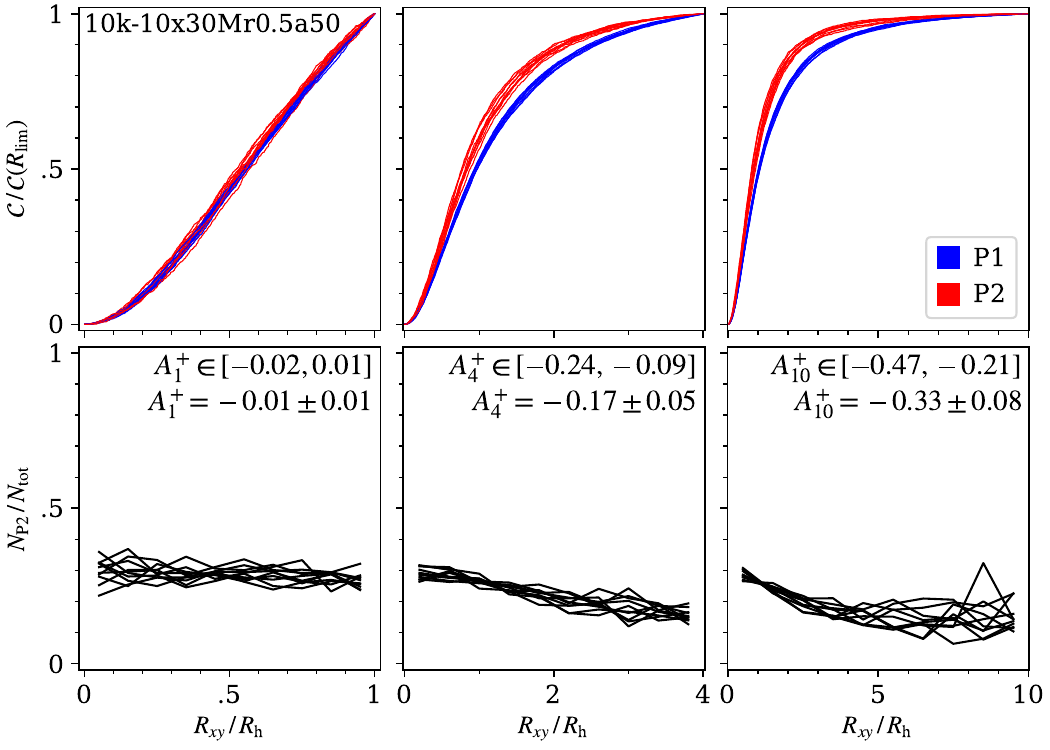}

    \caption{Selected \texttt{10k-*} models with the highest $A^+$ at $4\,\trh$ for each primordial binary set-up (as labelled in the top left corners).
    Top panels: Normalised projected cumulative radial distributions of stars in two populations (P1 and P2 -- distinguished by colours) in the clusters, calculated from Eq.~\eqref{eq:Ap}. All realisations of the corresponding model are plotted. Columns correspond to different values of $\Rlim$.
    Bottom panels: The ratio of P2 stars in several radial bins \citep[similarly to][]{leitinger_etal23}. Each model realisation is plotted with a black line.
    The parameter $A^{+}_{R\mathrm{lim}}$ (its maximum range and median with standard deviation) is also displayed for each region.}
    \label{fig:Ap_10k_detail}
\end{figure*}
\begin{figure*}
    \centering
    \includegraphics[width=0.49\linewidth]{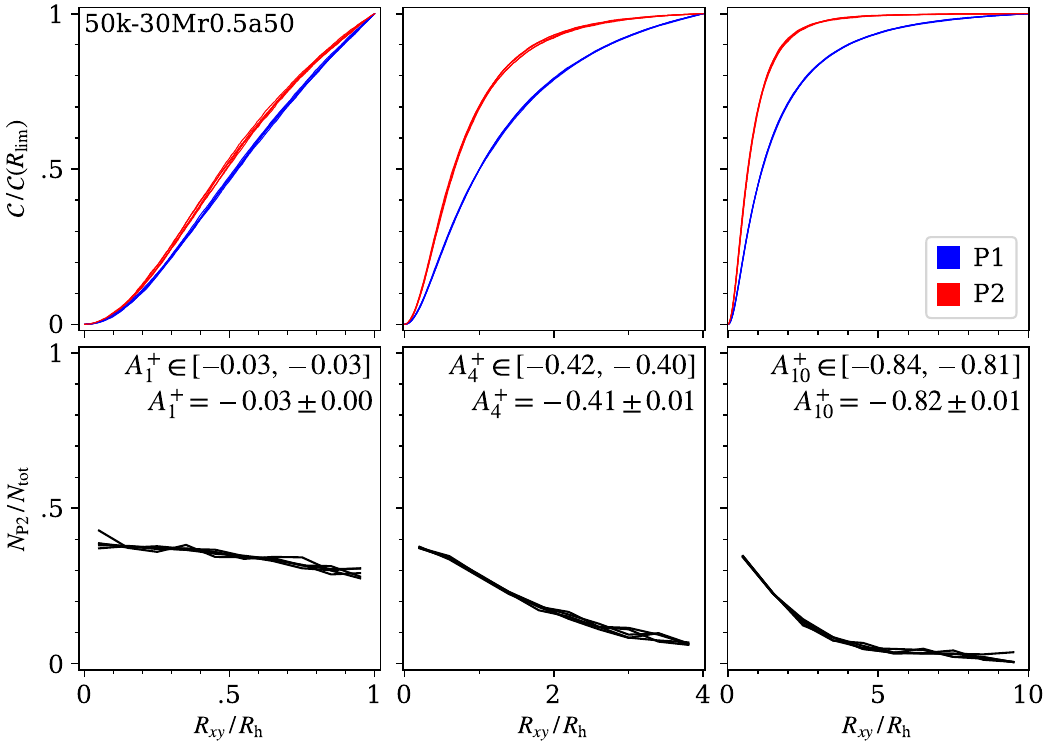}
    \hfill
    \includegraphics[width=0.49\linewidth]{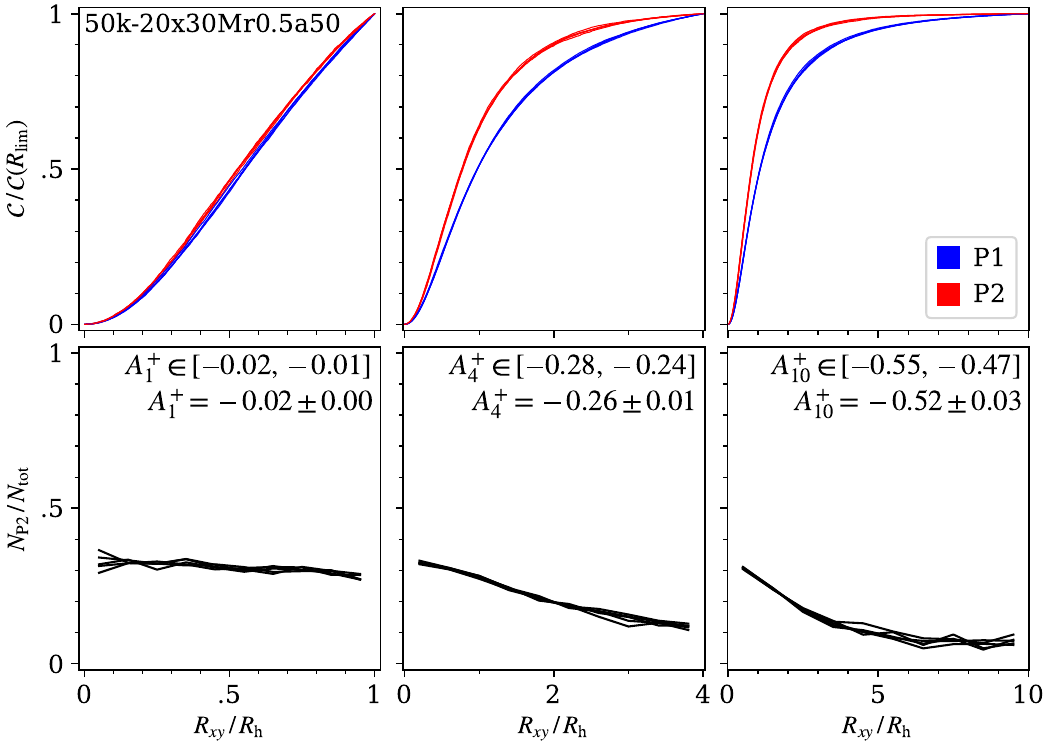}

    \caption{Same as Fig.~\ref{fig:Ap_10k_detail} but for the \texttt{50k-*} models.}
    \label{fig:Ap_50k_detail}
\end{figure*}

\begin{figure*}
    \centering
    \includegraphics[width=\linewidth]{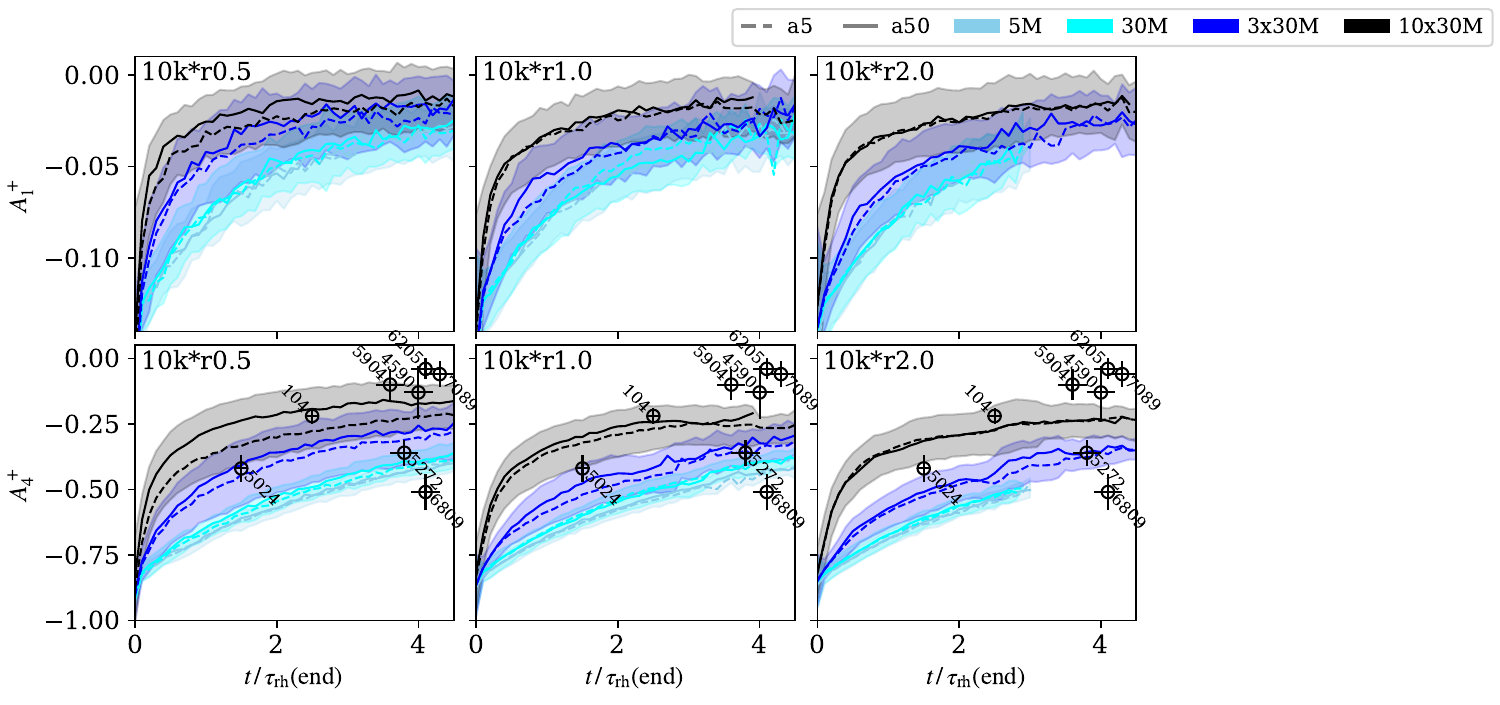}
    \caption{Time evolution of the parameter $A^{+}_{R\mathrm{lim}}$ in the \texttt{10k-*} clusters -- the top row is for $\Rlim = 1\,\rh$, the bottom one is for $\Rlim=4\,\rh$, see Eq.~\eqref{eq:Ap} for the definition. Only the models with the initial binary semi-major axis of 5\,au (dashed lines) and 50\,au (solid lines) are shown, averaged over all realisations of the models. The shaded area corresponds to their combined 1-sigma uncertainties. The $A^+_4$ data points of NGC clusters are taken from \citet{leitinger_etal24} for comparison.}
    \label{fig:Ap_10k_time}
\vspace{\floatsep} 
    \centering
    \includegraphics[width=\linewidth]{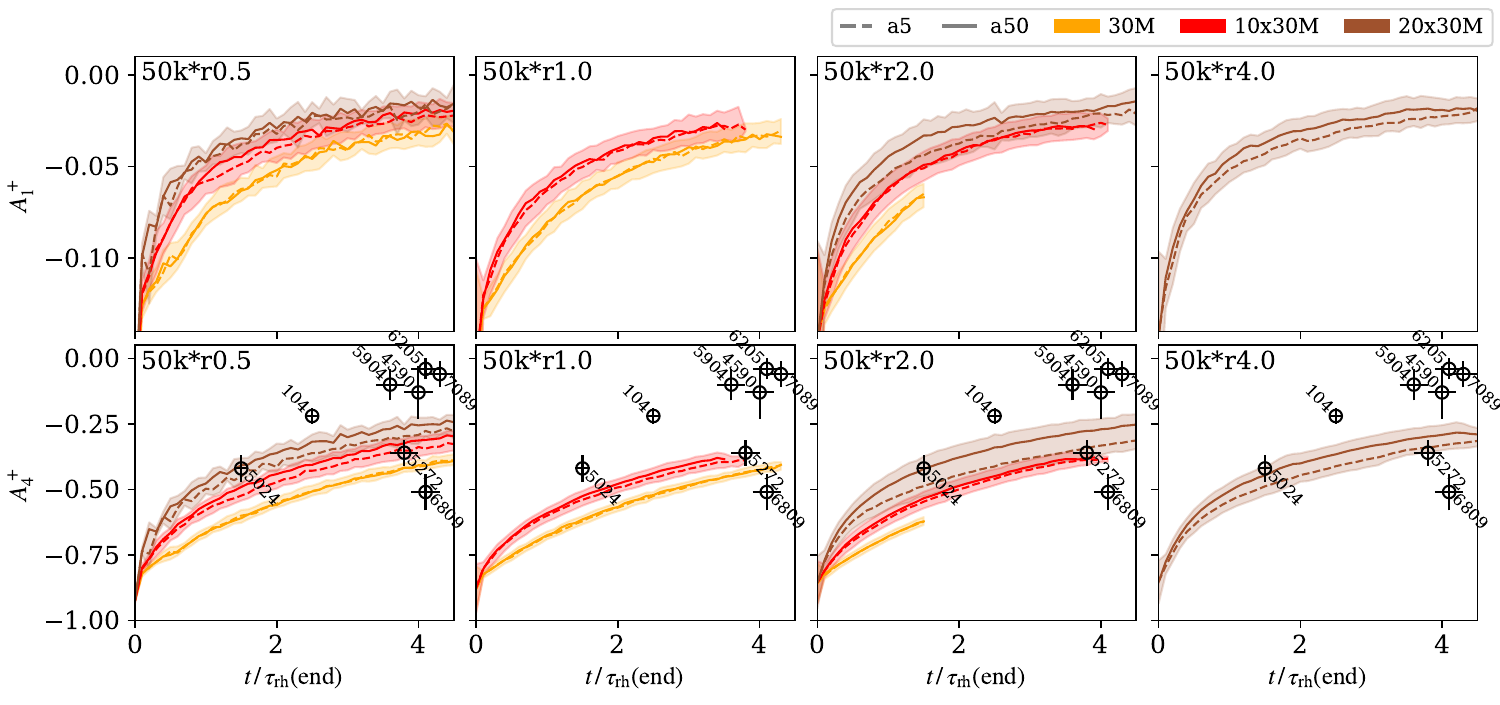}
    \caption{Same as Fig.~\ref{fig:Ap_10k_time} but for the \texttt{50k-*} models.}
    \label{fig:Ap_50k_time}
\end{figure*}

\section{Evolution of star clusters and binary stars}
\label{sec:results}

First, we explore the GCs' evolution, focusing on their 3D structure. The time evolution of the relaxation time is plotted in Fig.~\ref{fig:Trh}, calculated from Eq.~\eqref{eq:trh}. We note that although our models are isolated, we take into account that some stars might be escaping when calculating $\trh$. Thus, we iteratively determined the number of stars $N$ (and the enclosed cluster mass, $M_{\rm cl}$) within the cluster's Jacobi radius
\begin{equation}
    \label{eq:Rjac}
    r_{\rm J} = x_{\rm Gal} \left( \frac{\Mcl}{3 M_{\rm Gal}} \right)^{1/3} \,,
\end{equation}
with the assumed orbit of the cluster at $x_{\rm Gal} = 5\,\kpc$, corresponding to the Galactic mass of $M_{\rm Gal} \approx 5\cdot 10^{10}\,\Msun$ \citep{galaxy_mass, galaxy_mass_new}.
Fig.~\ref{fig:Trh} shows that the models with one $(5{+}5)\,\Msun$ or one $(30{+}30)\,\Msun$ binary (i.e.\ \texttt{10k-5M*} and \texttt{*-30M*}) have near-constant $\trh(t)$ and therefore near-constant $\rh(t)$. They evolve analogously to single-component models without binaries. In other words, the heating power of a single stellar-mass binary BH is either not high enough to cause a notable expansion of the cluster overall, or the $m_3$-stars do not go through as many scattering encounters on the same time scale as in the models with higher binary counts.

Similarly, we can infer from Fig.~\ref{fig:Trc} that the cores of the clusters with a single low-mass binary (i.e.~\texttt{10k-5M*} and \texttt{50k-30M*} models) gradually proceed towards core collapse. On the other hand, the cores of the models with massive or multiple binary stars contract more rapidly initially. In the case of the models \texttt{50k-10x30M*}, \texttt{50k-20x30M*}, and to an extent also \texttt{10k-10x30M}, Fig.~\ref{fig:Trc} also shows an increase of $\trc$ after the initial drop, corresponding to the expansion of the core due to binary heating.

Fig.~\ref{fig:Trh} further shows that $\trh(t)$ rises over time in the models with more massive binaries. This expansion is greater in the following scenarios:
\begin{enumerate}
    \item In the most compact clusters, because the relaxation processes are more rapid and the binary--single star interactions can occur more frequently in denser environments (compare the different line styles in each panel of Fig.~\ref{fig:Trh}).
    \item In the clusters with wider binaries, because the scattering cross-section of each binary is larger (e.g.~compare the line bifurcation in the models \texttt{10k-10x30Mr0.5*} in the fourth panel from the left of the figure).
    \item In the models with more binaries, as their scattering power increases (compare the same line styles across the panels of the figure).
\end{enumerate}
Consequently, the present-day $\trh$ of the GCs with multiple binaries does not correspond to the relaxation time scale at birth. This means that the dynamical age of such GCs might be underestimated if only inferred from their present-day structure. For comparison, we present some of our results (e.g.~in Figs.~\ref{fig:Nbin_10k} and~\ref{fig:Nbin_50k}) in terms of the initial and present-day (i.e.\ end-of-simulation) $\trh$ simultaneously. And later, when we compare our models with observations, we do so only in terms of the present-day $\trh$.

We must also emphasise that while the dynamical heating increases with the number of primordial binaries in the GC, it is not exactly proportional to it because these binaries also interact with each other and destroy themselves. We show this in Figs.~\ref{fig:Nbin_10k} and~\ref{fig:Nbin_50k}, where we plot the number of massive binaries in our models over time. For instance, see the panels for \texttt{10k-30M*} and \texttt{10k-3x30M*} in Fig.~\ref{fig:Nbin_10k} -- the latter had three binaries initially, however, most of them are quickly destroyed and, on average, only one or two remain after $1\,\trh$, while the former started and finished the simulation with one binary. This binary--binary destruction is more prevalent when the binaries have larger semi-major axes -- see that in most cases, the lighter-coloured lines (tighter binaries) are above the darker-coloured lines (wider binaries) in these figures.
Temporary oscillations of $\Nbin$ in Figs.~\ref{fig:Nbin_10k} and~\ref{fig:Nbin_50k} are mainly due to the binary capturing a single star and entering a triple system (occasionally also a quadruple), and we do not count multiples in these figures.

Comparing the plots in Fig.~\ref{fig:Trh} with the highest number of primordial binaries (i.e.~\texttt{10k-10x30M*} and \texttt{50k-20x30M*}), we see that even with twice as many binaries, $\trh(t)$ does not grow as much in the latter model. This is due to the smaller mass ratio $\Nbin(m_1{+}m_2)/\Mcl$ in the \texttt{50k-20x30M*} models than in the \texttt{10k-10x30M*} models. Moreover, this ratio is further lowered as the number of binary BHs in the \texttt{50k-20x30M*} models drops to a number between four and ten (see Fig.~\ref{fig:Nbin_50k}), comparable to the final binary count in \texttt{10k-10x30M*} (see Fig.~\ref{fig:Nbin_10k}). A much more numerous BH subsystem or more massive BHs might be necessary to cause an expansion of \texttt{50k-20x30M*} similar to the \texttt{10k-10x30M*} model.

Moreover, Fig.~\ref{fig:Trc} further shows that the ratio $\trc/\trh$ in the \texttt{50k-20x30M*} models is higher than in the \texttt{10k-10x30M*} models. This signals (as discussed in Sec.~\ref{sec:theory}) that the binaries in the \texttt{50k-20x30M*} models cannot interact with as many core P2 stars as in the \texttt{10k-10x30M*} models, and the population mixing in the former models would be less effective.

In Figs.~\ref{fig:alpha_10k} and~\ref{fig:alpha_50k}, we show the distribution of $\alpha$, introduced in Sect.~\ref{sec:theory} as the change of the binary binding energy per interaction with a single star.
We also plot the theoretical lower limit for the interactions that should lead to the escape of the $m_3$-stars.
Comparing the models of different initial sizes (panels on the same lines in the figures), we see that as $\rh(0)$ of the system increases, the probability of strong interactions leading to escape drops.
This is most visible in the models with a single massive binary star (i.e.~\texttt{10k-30M*} in Fig.~\ref{fig:alpha_10k} or \texttt{50k-30M*} in Fig.~\ref{fig:alpha_50k}).
While even the wider models with $(30{+}30)\,\Msun$ binaries still produce a population of escaping stars, we may notice that the \texttt{10k-5Mr2.0*} models with the least massive primordial binary are essentially unable to eject stars (see the top-right panel of Fig.~\ref{fig:alpha_10k}), thus any population inversion or mixing in these models should happen at the lowest rate.

While the models with multiple binaries also produce a population of escapers, their distributions of $\alpha$ are more shifted below the limiting value for escape (compare the models in the same column in Figs.~\ref{fig:alpha_10k} or~\ref{fig:alpha_50k}, e.g.\ \texttt{50k-*r0.5*} in the left-hand panels of Fig.~\ref{fig:alpha_50k}). This suggests that as the binary star number in the system increases, a higher number of the single $m_3$-stars that pass through the core region are subject to low-energy binary--single interactions rather than a single encounter that would kick them out of the cluster.
Consequently, we may expect that the binary--single interactions leading to the mixing (or inversion) of the radial distributions of P1 and P2 would predominantly happen in the more compact clusters with the more massive binary stars. Or in other words, that the more compact clusters would mix more rapidly than the extended clusters of the same mass and binaries.

\section{Comparison with observations}
\label{sec:observations}

To evaluate the effect of population redistribution due to binary heating, we calculate the area parameter \citep[introduced by][]{alessandrini_etal16} which shows the relative abundance of P1 and P2 from the projected normalised cumulative radial distributions in several radial regions:
\begin{equation}
    \label{eq:Ap}
    A^{+}_{R\mathrm{lim}}
    = \frac{1}{\Rh}
      \int_0^{\Rlim}{\left[ \frac{\CRD_\mathrm{P1}(R)}{\CRD_{\rm P1}(\Rlim)} -
                            \frac{\CRD_\mathrm{P2}(R)}{\CRD_{\rm P2}(\Rlim)} \right] \der R} \,,
\end{equation}
where $\CRD(R)$ is the cumulative radial distribution of stars at a projected radius $R$ for the specified population, and $\CRD(\Rlim)$ ensures its normalisation to one at the limiting radius. Hence, the methodology is similar to \citet[][their Figure~14 and Appendix~A]{leitinger_etal23}. For our analysis, we assume $\Rlim = 1\,\Rh$ (central region), $4\,\Rh$ \citep[to compare with][]{leitinger_etal23,leitinger_etal24}, and also $10\,\Rh$ (for the whole cluster), where $\Rh$ is the projected half-mass radius.

In Figs.~\ref{fig:Ap_10k_detail} and~\ref{fig:Ap_50k_detail}, we plot the radial distributions of P1 and P2 stars in selected models at $4\,\trh$ (each with a different number of binaries) where we found the highest values of $A^+$. All these simulations correspond to $\rh(0)=0.5\,\pc$, which is expected because the densest clusters are also the most dynamically active, and the binary--single star scattering there is more effective.
In all models, P1 and P2 mix in the central regions up to the half-mass radius (i.e.~$A_1^+ \approx 0$, albeit the values are slightly more negative in the models with fewer or less massive binaries).
The cumulative distribution functions of the model \texttt{10k-10x30Mr0.5a50} in Fig.~\ref{fig:Ap_10k_detail} show a higher level of mixing even up to $4$ or $10\,\Rh$, proving that a higher number of binary stars can mix the central and outer populations more than fewer binaries on the same dynamical time scale.
This is further supported by the bottom panels of each set of plots in the same figure, which show that the depletion of P2 stars from the core and their migration to the outer regions is dominant in the models with multiple binaries.

The more populous \texttt{50k-*} models plotted in Fig.~\ref{fig:Ap_50k_detail} do not exhibit as much variation between realisations as the less populous models in Fig.~\ref{fig:Ap_10k_detail} because the distributions are averaged over a larger number of stars in the clusters. On the other hand, the distributions of P2 stars in the bottom panels of Fig.~\ref{fig:Ap_50k_detail} show that the mixing power of even 20 primordial binaries with $30\,\Msun$ components is not enough to cause significant migration of the P2 stars outwards.

Furthermore, in Figs.~\ref{fig:Ap_10k_time} and~\ref{fig:Ap_50k_time}, we plot the time evolution of the parameter $A^+$ in the \texttt{10k-*} and \texttt{50k-*} clusters, respectively. All models have highly negative initial conditions (P2 is centrally concentrated by definition) and then gradually evolve towards a mixed state ($A^{+}{\approx}0$). We see that the more concentrated clusters and those with more binaries mix slightly more rapidly; however, all clusters with a given number of primordial binaries reach similar $A^+_1$ values at $1{-}2\,\trh$.
On the other hand, the more compact clusters show higher $A^+_4$ at any given time (compare the lines of the same colour across different panels of the figure). Similarly, the models with a lower binary count require significantly longer dynamical evolution to reach the same $A^+_4$ values as the more compact ones (compare the differently coloured lines in the same panel).\!\footnote{We note that while in the work of \citet{modest25_poster}, we showed almost full radial mixing of the clusters with $10^4$ stars and ten massive primordial binaries, those initial conditions corresponded to a very idealised toy model. In the models presented here, which have more realistic velocity distributions, the mixing caused by the binary is slightly suppressed by the random radial motions of the single stars.}
We note that our models does not showcase a clear population inversion ($A_1^+ \gg 0$) anywhere up to $4.5\,\trh$.

In the bottom panels of both figures, we plot several dynamically-young GCs from the sample of \citet{leitinger_etal23}. Focusing only on the \texttt{10k-*} models in Fig.~\ref{fig:Ap_10k_time}, we see that those with ten primordial binaries and $\rh(0)=0.5\,\pc$ fit several of the very mixed GCs within their 1-sigma uncertainties. The less-mixed GCs (e.g.\ NGC~5024, 5272, and 6809) are, in turn, well represented by the models with fewer primordial binaries or the more expanded clusters.

Comparing Fig.~\ref{fig:Ap_10k_time} and~\ref{fig:Ap_50k_time}, the \texttt{50k-*} models show lower spread of $A^+$ than the less populous \texttt{10k-*} models. These fluctuations and inconsistencies between realisations of the same model are caused by the specific dynamical evolution and binary--binary and binary--single star interactions in each cluster, which either lead to their escape or ionisation, temporarily reducing the binary heating engine in the core (as we discussed also in terms of the radial distribution of P2 stars). We also see that the mixing of P1 and P2 is not as effective in the more populous clusters, even in the most compact ones with 20 primordial binaries. On the other hand, the trend we described above (`more binaries equal more mixing') is still visible. This suggests that in a more populous cluster, either a more numerous stellar-mass BH--BH subsystem is necessary, or we need to increase the BH masses -- to a degree, this is also suggested by the theoretical derivations we did in Sec.~\ref{sec:theory}, specifically in Fig.~\ref{fig:estimates}.

While our results do not provide conclusive evidence for the cause of P1/P2 mixing in young GCs, they are highly suggestive that the dynamical heating by binaries plays a role, especially when it is set appropriately to the size of the system. It is, therefore, worth investigating this in subsequent studies.

\section{Discussion}
\label{sec:discussion}

We acknowledge that we have only explored numerically the domain of clusters with ten to fifty thousand stars, and we need to map those results to `real' GCs containing up to a few million particles. The distributions of $\alpha$ (presented in Figs.~\ref{fig:alpha_10k} and~\ref{fig:alpha_50k}) show that the results on binary--single star scatterings leading to stellar escapes in our models are comparable between the models, although one has five times more stars (e.g.~compare the curves for \texttt{10k-30M*} and \texttt{50k-30M*}).
This also refers back to Fig.~\ref{fig:estimates}, where we see that the relationship between the binary semi-major axis and the star cluster size and mass is self-similar. Therefore, this knowledge about the distribution of $\alpha$ from our less populous models can be applied to understanding what fraction of encounters would lead to the ejection from the core, the ejection from the cluster entirely, or placing the core stars in the halo in real GCs.

Despite our designing of star clusters where massive binary stars could efficiently eject the central population of P2 stars, our models do not show the radial inversion of P1 and P2 but only mixing. Nonetheless, we must note that out of the 13~dynamically-young GCs in the sample of \citet{leitinger_etal23, leitinger_etal24}, four have $A^+_4 < 0$, two have $A^+_4 > 0$, and seven have $A^+_4 \approx 0$. This indicates -- although the sample is small -- that high positive $A^+$ values may be rare to achieve.
We also note that the two clusters with $A^+_4 > 0$ have the following radial distributions. NGC~3201 primarily contains P2 stars within the range up to $0.5\,\Rh$ and beyond $1.5\,\Rh$, and P1 stars in the range from $0.5\,\Rh$ to $1.5\,\Rh$. On the other hand, NGC~6101 shows a consistent central concentration of P1 stars throughout the GC and is, therefore, technically the only GC with a clear P1 central concentration. Meanwhile, NGC 3201 may be indicative of a GC that began with a P2 central concentration and underwent a semi-successful inversion of its populations. Furthermore, both of these GCs with high positive $A^+_4$ are reportedly accreted from the Sequoia dwarf galaxy \citep{massari_etal19}, i.e.\ the formation conditions might have determined their structural properties more than in the other GCs \citep[see also Figure 15 in][which connects the progenitor system of the GCs]{leitinger_etal24}.

More to the point, while the binary heating engine in our models caused the outward migration of the central P2 stars, it also pushed all the sinking P1 stars to higher radii as the cluster relaxed. We introduced four evolutionary time scales in Sec.~\ref{sec:theory} -- those are
1) the time scale for individual encounters between a given binary and a random single star,
2) the time scale to eject a significant amount of P2 through binary--single star encounters from the core,
3) the time scale to sufficiently mix up the stars in the core, such that all core stars have the opportunity to be ejected, which is related to the core relaxation time
and 4) the time scale for the infall of P1, which is related to the relaxation time scale of the entire system.
To achieve the radial inversion of P1 and P2, the ejection of P2 from the core must be more rapid than the infall of P1, and once most of the P2 stars are gone from the core, it has to stop before P1 builds up in the core. If, however, the binary engine continues to scatter, or the infall of P1 happens on a shorter time scale than the ejection of the core stars, the only possible outcome is P1/P2 mixing.
Although we saw in our models that any high initial binary count has been reduced over time it was never fully stopped. The kicking of stellar-mass BHs out is rather stochastic and cannot be timed or predicted from the initial conditions. Consequently, all our models mixed P1 and P2 in the long term.

An alternative mechanism to quench the binary--single star scatterings is to have a single binary system composed of two intermediate-mass black holes (IMBHs).
Let us assume one of our modelled clusters with $N=50\,000$ stars (where we take $0.1N$ as the core P2 stars), $\rh[,0]=0.5\,\pc$, $m_3=1\,\Msun$, and one BH--BH binary of $m_1=m_2=30\,\Msun$. Then the ejection time scale, according to Eq.~\eqref{eq:tej} and the typical value of $\alpha \approx 10^{-1.8}$ (see Fig.~\ref{fig:alpha_50k}), is $\tej \approx 1.8 \cdot 10^3\,\Myr$. This is very long compared to $\trh \approx 16.5\,\Myr$ and $\trc \approx 8\,\Myr$ (see the \texttt{50k-30M*} panels in Figs.~\ref{fig:Trh} and~\ref{fig:Trc}). However, when we replace the stellar-mass BH with an IMBH of $m_1=m_2=300\,\Msun$, the encounter time scale from Eq.~\eqref{eq:tenc} becomes a thousand times smaller and $\tej \approx 1.8\,\Myr$. Comparing this with the \texttt{50k-10x30M*} panels in Figs.~\ref{fig:Trh} and~\ref{fig:Trc}, where the mass contained in the stellar mass BHs is equivalent to this IMBH--IMBH binary, we get $\trc\approx\tej\ll\trh$. Therefore, such IMBHs could eject the P2 stars as effectively as ten stellar-mass BH binaries.
Unlike stellar-mass BHs, the IMBHs could also rapidly merge due to gravitational wave emission and leave the GC core due to the recoil kick \citep{Campanelli_etal_2007, Gonzalez_etal_2007} that can have up to thousands of km/s \citep[e.g.][]{Mahapatra_etal_2024}.
The scattering of P2 stars would also make the IMBH binary more eccentric, helping with merging and further amplifying the kick \citep{Sperhake_etal_2020}.
For context, the general relativistic merger time of an eccentric BH--BH binary is \citep[see, e.g.][]{davies_etal2011}
\begin{equation}
    \label{eq:tgr}
    \tgr \approx 1.825 \cdot 10^{12}\,\Myr \
        \left( \frac{\abin}{\au} \right)^4
        \frac{\Msun^3}{m_1 m_2 (m_1 + m_2)}
        \left( 1-e^2 \right)^{7/2} .
\end{equation}

We also refer to \citet{mpop_mocca_dynamics}, who argue that the change in the external tidal field and tidal stripping may be one of the main reasons for the P1/P2 radial inversion, albeit for all of their simulated clusters, such an inversion was only marginal and transient.
Another mechanism that would lower the central potential well and push the central P2 population outward is gas expulsion \citep[e.g.][]{decressin_etal2010}. However, gas loss would affect all stars in the GC. Thus, it is to be seen whether it could cause clear radial inversion of P1 and P2, since the scattering of P2 needs to be more localised in the centre.
Lastly, we note that while it is conventionally assumed that P2 is born in the centres of GCs, there is still a possibility for another formation scenario of P2.
All these additional aspects that could lead to faster mixing or the radial inversion of P1 and P2, which we were not able to explore in their entirety in our simple dynamical scenario, warrant future work.

\section{Conclusions}
\label{sec:conclusions}

We studied the dynamics of stars inside star clusters to understand why certain observed, dynamically-young GCs have their P1 more centrally concentrated than P2, some vice versa, and some GCs appear fully mixed.
Using theoretical arguments and their testing in specifically designed numerical simulations of various clusters, we found that massive binary stars may be responsible for mixing of P1 and P2 early in the GC evolution. They heat the cluster core through binary--single star interactions and facilitate population redistribution on a shorter time scale than relaxation processes.

First, we note that while those high-mass binaries can push the centrally born P2 stars outwards, we do not observe a clear radial inversion of the P1 and P2 stars in any of our numerical setups. Consequently, we still cannot explain the nature of systems such as NGC~3201 and NGC~6101, where P1 is more centrally concentrated than P2.
Provided, however, that only these two examples of inverted GCs are known to date and that they originated in a galaxy accreted by the Milky Way \citep{massari_etal19, leitinger_etal24}, this might suggest some specific evolution, such as tidal stripping \citep[see, e.g.][]{mpop_mocca_dynamics}, which is currently beyond the scope of this paper.

More importantly, we conclude with our numerical analysis that high-mass primordial binaries can mix the GC members up to the half-mass radius and, in more compact systems, even to higher radii.
While one massive binary is already able to affect the radial profile of the centrally-born P2 stars, full mixing of the populations is more effective if the models include several (e.g.~ten or twenty) primordial binaries. It is also more rapid in the most compact models with the initial $\rh(0)=0.5\,\pc$ rather than those with $\rh(0) = 1$, $2$ or $4\,\pc$. Hence, the radial density profile of P1 and P2 is affected the most 1) in denser GCs, 2) by more massive binaries, and 3) by a higher binary count. The mechanism proposed in this paper can help us explain the observations of some dynamically-young GCs, such as NGC~4590 or NGC~5904.

\begin{acknowledgements}
This research used computational resources from e-INFRA CZ (project ID:90254), supported by the Ministry of Education, Youth and Sports of the Czech Republic; and the computational cluster Virgo at the Astronomical Institute of the Czech Academy of Sciences.
V.P.~is funded by the European Union's Horizon Europe and the Central Bohemian Region under the Marie Skłodowska-Curie Actions -- COFUND, Grant agreement \href{https://doi.org/10.3030/101081195}{ID:101081195} (``MERIT''). V.P.~also acknowledges support from the project RVO:67985815 at the Czech Academy of Sciences.
E.I.L.~acknowledges support from the ERC Consolidator Grant funding scheme (project ASTEROCHRONOMETRY, \url{https://www.asterochronometry.eu}, G.A.~n.~772293).
A.B.~acknowledges support from the Australian Research Council (ARC) Centre of Excellence for Gravitational Wave Discovery (OzGrav), through project number CE230100016.
M.H.~acknowledges financial support from the Excellence Cluster ORIGINS which is funded by the Deutsche Forschungsgemeinschaft (DFG, German Research Foundation) under Germany’s Excellence Strategy – EXC 2094 – 390783311.
A.J.W.~has received funding from the European Union’s Horizon 2020 research and innovation programme under the Marie Skłodowska-Curie grant agreement No 101104656 and the Royal Society's University Research Fellowship, reference URF/R1/241791.
Last but not least, we thank the referee for valuable suggestions that improved this paper, in particular for pointing out the importance of the relative values of the core and half-mass relaxation time scales together with the time scale to eject stars from the core.
\end{acknowledgements}

\bibliographystyle{aa} 
\bibliography{main}


\end{document}